# Visualization Onboarding Grounded in Educational Theories

Christina Stoiber, Markus Wagner, Florian Grassinger, Margit Pohl, Holger Stitz, Marc Streit, Benjamin Potzmann and Wolfgang Aigner

**Abstract** The aim of visualization is to support people in dealing with large and complex information structures, to make these structures more comprehensible, facilitate exploration, and enable knowledge discovery. However, users often have problems reading and interpreting data from visualizations, in particular when they experience them for the first time. A lack of visualization literacy, i.e., knowledge in terms of domain, data, visual encoding, interaction, and also analytical methods can be observed. To support users in learning how to use new digital technologies, the

---

Christina Stoiber
St. Poelten University of Applied Sciences, Matthias Corvinus-Straße 15, 3100 St. Poelten, Austria, e-mail: `christina.stoiber@fhstp.ac.at`

Markus Wagner
St. Poelten University of Applied Sciences, Matthias Corvinus-Straße 15, 3100 St. Poelten, Austria, e-mail: `markus.wagner@fhstp.ac.at`

Florian Grassinger
St. Poelten University of Applied Sciences, Matthias Corvinus-Straße 15, 3100 St. Poelten, Austria, e-mail: `florian.grassinger@fhstp.ac.at`

Margit Pohl
TU Wien, Argentinierstrasse 8, 1040 Vienna, Austria e-mail: `margit.pohl@tuwien.ac.at`

Holger Stitz
datavisyn GmbH, Altenberger Straße 69 4040 Linz, Austria, e-mail: `holger.stitz@datavisyn.io`

Marc Streit
Johannes Kepler University, Altenberger Straße 69, 4040 Linz, Austria, e-mail: `marc.streit@jku.at`

Benjamin Potzmann
TU Wien, Argentinierstrasse 8, 1040 Vienna, Austria e-mail: `benjamin.potzmann@tuwien.ac.at`

Wolfgang Aigner
St. Poelten University of Applied Sciences, Matthias Corvinus-Straße 15, 3100 St. Poelten, Austria e-mail: `wolfgang.aigner@fhstp.ac.at`

concept of onboarding has been successfully applied in other domains. However, it has not received much attention from the visualization community so far. This chapter aims to fill this gap by defining the concept and systematically laying out the design space of onboarding in the context of visualization as a descriptive design space. On this basis, we present a survey of approaches from the academic community as well as from commercial products, especially surveying educational theories that inform the onboarding strategies. Additionally, we derived design considerations based on previous publications and present some guidelines for the design of visualization onboarding concepts.

# 1 Introduction

The term onboarding was originally coined in the context of human resources management to support new employees in learning about their tasks that are part of their job within a particular company [37]. The aim of this ongoing process is to communicate formal knowledge about their tasks, but also informal knowledge about organizational culture and its unwritten rules, to the new employees. This concept has been transferred to other domains such as human-computer interaction (HCI) [9, 46, 7, 35, 17, 5, 21]. More recently, the focus of onboarding has shifted towards mobile applications. Hulik[1] introduced the concept of supporting users in learning smartphone applications and software tools. Kumar defined *user onboarding* as "the process of increasing the likelihood that new users become successful when adopting your product." [39].

We think that it is also useful to conceptualize the process of learning about complex visualizations that cannot be understood at a first glance by having visualization onboarding concepts. We define visualization onboarding as the following: ***"Visualization onboarding is the process of supporting users in reading, interpreting, and extracting information from visual representations of data."*** [59] This learning process often takes place immediately before or while users work with the visualization and is highly task-oriented. In this context, theories about learning play an important role. In the visualization community, a considerable amount of research has addressed the question how to increase visualization literacy (see e.g., [51], [30]). This research is generally based on educational theories from psychology, especially on constructivist research [15]. The basic assumption is that knowledge about visualizations can best be acquired by creating one's own visualization and actively generating one's own view about this topic. Similarly, educational theories can also be adopted to explain the usefulness of onboarding approaches.

In the literature, several different possibilities how to realize onboarding have been suggested (see Table 1). Some of them are primarily based on cognitivist approaches (e.g., tutorials) [45, 40] or Gestalt psychology (using analogy as a learning principle [52]). The educational theory on which these solutions are based are some-times reflected explicitly and sometimes not. The discussion of this topic could help to clarify which approaches in the design of onboarding systems are more helpful than others. Informal evidence indicates that tutorials are often not read, and users just proceed and start working and exploring features of the system themselves. Nevertheless, commercial systems often rely on tutorials as well as help websites as onboarding systems e.g., [61, 43, 1, 31].

---

[1] https://useronboard.com, Accessed: 2021-04-30

In this book chapter, we present a descriptive design space, presented in Figure 2, covering aspects of visualization onboarding especially with the focus on educational theories. We conduct a systematic literature review to identify the state of the art in visualization onboarding and to categorize the work by summarizing existing onboarding concepts in scientific publications and commercial visualization tools using the *Five W's and How* [23, 24]. **WHY** is visualization onboarding needed? **WHAT** is visualization onboarding? **WHO** is the user? Which knowledge gap does the user have? **HOW** is visualization onboarding provided? **WHERE** is visualization onboarding provided? **WHEN** is visualization onboarding used? Additionally, we derived design considerations based on the collected publications and provide some existing guidelines for the application of educational theories for visualization onboarding in Section 4.4. Overall, we can report that whether other approaches are better for onboarding or not is still an open question,. Empirical research based on educational theories could help to gain more systematic information about this area.

## 2 Related Work

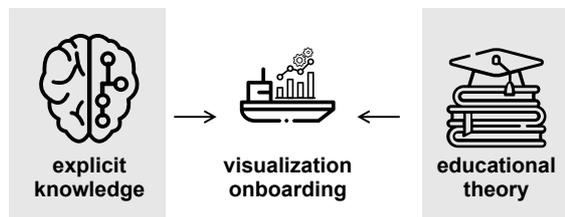

**Fig. 1** Visualization onboarding aims to support end users in comprehending data visualizations and take full advantage of the tools at hand. With effectively designed onboarding methods the knowledge gap of users could be filled. Thus, it makes sense to tap in the field of educational theories as well as identify how onboarding can benefit from explicit knowledge.

As visualization onboarding aims at filling the knowledge gaps of users by supporting the learning of new concepts, it makes sense to build upon knowledge from the fields of learning theories and cognitive science (see Figure 1). Therefore, we present the related work for visualization onboarding, educational theories in visualization and cognitive science, as well as how explicit knowledge relates to onboarding in the following subsections.

## 2.1 Visualization Onboarding

So far, there has been little discussion about onboarding concepts for visualization techniques and visual analytics (VA) tools. Tanahashi et al. [62] investigated top-down and bottom-up teaching methods as well as active or passive learning types. The bottom-up teaching method ("textbook approach") [70] focuses on small, detailed pieces of information which students then combine to get a better understanding. Besides, a top-down teaching method is given when a broad overview first helps to understand the abstract, high-level parts of an idea/topic which then provide context for understanding its components in detail [62]. Furthermore, a distinction can be made between active and passive learning types. Passive learning means that students only receive the information without participatory dialog. In contrast, active learning describes an active participation [62]. Their analysis indicates that top-down exercises were more effective than bottom-up and active learning types with top-down tasks the most effective ones. In their comparative study, Kwon and Lee [40] explored the effectiveness of active learning strategies. Three tutorial types—static, video-based, and interactive (see Figure 4 (3))—were used to support the learning of parallel coordinates plot visualizations. They observed that participants who used interactive and video tutorials outperformed participants who used static or no tutorials at all. In a study which set out to determine the power of teaching unfamiliar visualization by linking it to a more familiar one, Ruchikachorn and Mueller [52] found out that the learning by analogy concept is useful as participants in their study could understand the unfamiliar visualization methods fully or at least significantly better after they observed or interacted with the transitions from the familiar counterpart. They assessed four combinations and compared their difference in visual literacy: scatter plot matrix against hyperbox, linear chart against spiral chart (see Figure 4 (2)), hierarchical pie chart against treemap, and data table against parallel coordinates plots. The authors describe also another advantage of learning-by-analogy over other forms of demonstrations such as textual or oral descriptions is the power of visuals, as they bridge any language barriers. The educational community has also studied how students interpret and generate data visualizations [4] and how to teach bar charts in early grades [2] using a tablet app, called "C'est la vis", supporting elementary school pupils to learn how to interpret bar charts based on the concreteness fading approach. Concreteness fading is a pedagogical method where new concepts are presented with concrete examples at first, before progressively abstracting them (see Figure 3 (2)). Recently, Bishop et al. [8] developed a tablet-based tool called Construct-A-Vis, which supports elementary school children in creating visualization based on free-form activities. They used scaffolding as a pedagogical method. Scaffolding defined as "the process that enables a child or novice to solve a problem, carry out a task, or achieve a goal which would be beyond his unassisted efforts" [68], issued in education to help children acquire new knowledge and skills. In detail, they integrated feedback mechanisms by showing if the visual mapping was correct. Additionally, Bishop et al. [18] developed an interactive pedagogical method for training and cognition of a treemap design, as well as a treemap literacy test, seen in Figure 3 (4). The user study showed that students who interacted with the teaching tool outperformed those students who learned through slides.

In addition to scientific literature, onboarding concepts are integrated in commercial visualization tools as well. Nowadays most of these commercial visualization tools already integrate onboarding concepts focusing on the explanation of features, see Table 1. Yalçin [69] presented HelpIn, a design of a contextual in-situ help system to explain features of Keshif [36]. Furthermore, IBM Cognos Analytics [31] (see Figure 3 (1)), for example, uses step-by-step tours with tooltips and overlays for onboarding new users. A more traditional approach is used by the commercial visualization tool Advizor [1] which makes use of textual descriptions to explain the visual mapping (seen in Figure 4 (1)) for visualization techniques.

Besides, there are platforms and websites available which can be categorized as external onboarding methods [59] supporting users in understanding the visual mapping of various visualization technqiues. For instance, *The graphic continuum* [60] provides an overview of visualization types and supports design and method decisions. Similarly, the *Data Visualisation Catalogue* [50] seeks to support users to understand the encoding and building blocks of different visualization types. Furthermore, *From Data to Viz* [27] aims to find an appropriate visualization type based on the input data using a decision tree. The catalogue offers definitions, variations, and the use of each visualization type in addition to potential issues that may arise during use and interpretation. These systems are not related to a particular visualization tool, neither integrate any educational theories. In recent literature, Wang et al. [66] presented a set of cheat sheets to support visualization literacy around visualization techniques inspired by infographics, data comics, and cheat sheets that are established onboarding methods in domains such as machine learning.

## 2.2 Educational Theories in Visualization and Cognitive Science

Visualization onboarding supports users in learning new concepts [59]; therefore it makes sense to build upon the knowledge from the field of learning theories and cognitive science. We distinguish between three main educational theories: behaviorism, cognitivism, and constructivism [15]. Behaviorism is an educational theory that only focuses on objectively observable behaviors and discounts any independent activities of the mind [67]. It is based on positive and negative reinforcement techniques. Besides, Cognitivism is a philosophy of learning, founded on the premise that learning can be modelled as a kind of information processing [15]. Each of us generates our own "rules" and "mental models," which we use to make sense of our experiences. Learning, therefore, is the process of adjusting our mental models to accommodate new experiences. E-learning systems often integrate elements from different educational theories. This also applies to most onboarding systems.

Constructivist theories seem to be the one most appropriate for explaining learning processes with onboarding systems because they reflect on the application of learning in a practical context. The concept of cognitive apprenticeship plays an important role in constructivism [56, 14]. *Cognitive apprenticeship* is a kind of guided participation by learners in real processes of knowledge generation. This is related to the concept of *scaffolding* [26] where teachers gradually reduce the level of support for the student until the student is able to work autonomously. Cognitive apprenticeship and scaffolding can explain processes related to onboarding because the goal of the learners is to solve a real task, while the guidance is gradually reduced.

Another theoretical framework relevant for onboarding is *graph comprehension*, a theory that aims to explain how users make sense of graphs. Most of the investigations in this context deal with simple, small graphs [48]. Nevertheless, the findings from graph comprehension yield interesting results that can inform the design of visualizations. This is especially valuable for onboarding systems because investigations in this area often address the issue of how to design graphs that are appropriate for use in educational contexts. One of the most influential models in the context of the theory of graph comprehension describes this activity as consisting of three stages [20]. These three stages are: (1) reading the data (i.e., finding individual data values), (2) reading between the data (i.e., finding relationships between the data), (3) going beyond the data (i.e., interpreting the data, developing hypotheses about the data). Educational graphs are supposed to support all three stages but the ultimate goal is to induce learners to "go beyond the data", that is, to reflect on the data and draw conclusions. Shah et al. [58] argue that inexperienced users typically concentrate on single data points or single lines in line graphs, whereas experts are able to actually interpret patterns in the data. Peeck [47] investigated whether it is possible to motivate learners to process graphs more comprehensively. In this context, the author successfully tested whether specific instructions for the processing of graphs support learning. The author also postulates that other measures such as cues to draw the learner's attention or motivating the learner to solve simple tasks by using the graphs are beneficial. Based on this approach, it can be recommended that onboarding should especially support "going beyond the data" and that instructions and visual cues can help users to better understand visualizations.

A further learning theory relevant for onboarding is *Microlearning*. Microlearning as an approach is a reaction to several technical developments. First, mobile technologies enable learners to learn flexibly, e.g., on the way to work, while traveling on public transport or while waiting for a physician. In addition, microlearning is also relevant for workplace learning and continuing education [57]. Employees in companies or other organizations do not need lengthy explanations but focused information that is necessary to continue their work. Microlearning has been defined as "special moments or episodes of learning while dealing with specific tasks or content, and engaging in small but conscious steps." [29]. Microlearning usually encompasses small units of learning that never take longer than 15 minutes. The situation described for Microlearning in the context of workplace learning is similar to the situation of users of complex information visualization systems.

Finally, *Gestalt psychology* is a theory that might be relevant for the design of onboarding systems. It is well-known that Gestalt psychology has made important contributions in the area of perceptual psychology. It is less well-known that Gestalt psychologists also conducted relevant research in the area of reasoning and problem-solving (see, e.g., [42], [28]). This is especially interesting for the design of visualizations as Gestalt psychologists conceptualized problem-solving as the (sudden) perception of structure in a problem domain. The so-called "Aha"-moment is the moment when pieces fall into place and coherent structure is identified. In this context, the usage of analogies plays an important role because the transfer of structural knowledge from a well-known domain to an unknown domain is one of the learning methods that was suggested by Gestalt psychologists. Analogies can also be used to support onboarding in improving the understanding of complex visualizations as shown in the concept by Ruchikachorn and Mueller [52].

## 2.3 Knowledge Integration for Onboarding

In this Section, we describe how user onboarding can benefit from *explicit knowledge* sources and contribute to generate new *(tacit) knowledge* and insights [16, 65].

In this work, we mainly consider *explicit knowledge* [65] as the source for providing onboarding. Usually, two types of *prior knowledge* are needed by a user to analyze data: *operational knowledge* (how to interact with the information visualization system), and *domain knowledge* (how to interpret the content) [10]. While a focus on usability and a perception- and cognition-aware design can alleviate the need for operational knowledge, domain knowledge cannot be easily replaced [10]. Stoiber et al. [59] further enhanced the levels of the users' *prior knowledge* for visualization onboarding based on the nested model [44] as: 1) *domain knowledge* (e.g. vocabulary and concepts); 2) *data knowledge* (understanding the particular datatype); 3) *visual encoding knowledge* (understanding the visual mapping); 4) *interaction knowledge* (for performing tasks and understand relations in the data); and 5) *analytical knowledge* (knowledge of different automated data analysis methods) – see Section 3.2.1 for more details. However, Chen et al. [10] as well as Stoiber et al. [59] described the term *prior knowledge* at different granularities, whereby *operational knowledge* [10] can be seen as similar to the combination of *visual encoding, interaction, and analytical knowledge* [59].

Based on the previously introduced terminology, we further characterize knowledge in Section 3.2.1, listing all possible knowledge types needed for supporting meaningful onboarding.

## 3 Descriptive Design Space

In our previous work [59], we introduced a descriptive design space for visualization onboarding. This work enhances the design space and discusses the role of educational theories in the context of onboarding.

### 3.1 Construction of Design Space

We structured the design space based on *Five W's* and the appended *How* [23, 24]. These questions are frequently used to describe a matter from its most relevant angles in technical documentation and communication. Furthermore, the same questions have already been employed for structuring the use of visualization for healthcare informatics [71] and in a survey on the role of visual analytics in deep learning research [25]. We describe the space of visualization onboarding along the following questions: **WHO** is the user? Which knowledge gaps does the user have? **HOW** is visualization onboarding provided? **WHERE** is visualization onboarding provided? **WHEN** is visualization onboarding used? Inside of each dimension (question), we defined several categories which are described in detail in the section below. We followed an open coding approach for the survey of onboarding concepts where we unified top-down approaches as well as bottom-up categorizations. Where available, we used existing taxonomies or frameworks, which we adapt to the specifics of visualization onboarding.

### 3.2 Design Space Dimensions

The aim of visualization onboarding is to support human in dealing with large and complex information structures, to make them more comprehensible, facilitate exploration, and enable knowledge discovery. Nevertheless, user often have problems in reading and interpreting data from visualizations, in particular when they experience them for the first time. In this section, we present the design space dimensions of visualization onboarding and show its various aspects.

#### 3.2.1 WHO is the user?

Users need to understand the process and reasoning that lead to the visual appearance, interactive behavior, and findings. Making this process transparent to the users is a central aspect in the design of visual analytics solutions. For conceptualizing this aspect, we adapt the nested model by Munzner and colleagues [44] as the guiding framework for presenting different levels of knowledge. The nested model is a unified approach that splits the design into four levels and combines these with appropriate evaluation methods to mitigate threats to validity at each level. In order to be able to cover visual analytic approaches and include automated data analysis components, we expand the original model by adding analytical methods alongside visual encoding/interaction idioms. Analytical knowledge–such as different automated data analysis approaches, machine learning methods, or statistical methods applied to the data–is necessary to understand complex visualization interfaces and data. Figure 2 (3) keeps the nesting but shows an altered representation of the different levels. The model components represent the different levels of knowledge that (a) visualization users need in order to correctly interpret (interactive) visualization artifacts and (b) visualization designers have to consider when developing onboarding concepts.

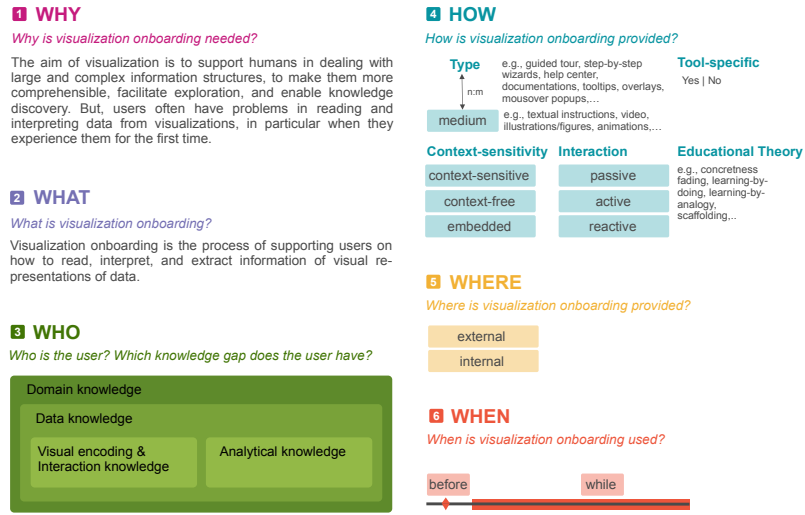

**Fig. 2** A visual overview of the onboarding design space and of how all six questions, "Why, What, Who, How, Where, and When" relate to one another. Each question corresponds to one paper section as indicated by the numbered tag near each question title.

Based on the definitions of our understanding of knowledge in the context of VA (see Section 2.3), we define the following levels: *domain knowledge, data knowledge, knowledge of visual encoding & interaction concepts, and analytical knowledge*.

**Domain knowledge**: A specific domain is a particular field of interest by target users of a visualization tool (e.g., medicine, data journalism, bioinformatics). Each domain has its own vocabulary for describing the data and problems, workflows, and how data can be used to solve a problem. Domain knowledge is also an ensemble of concepts, intellectual tools, and informational resources that a user can draw upon to put the visualized data into context.

**Data knowledge**: Many visualization tools are specific to a particular *type of data*, such as multivariate data, hierarchical data, network data, or time-oriented data. Data knowledge refers to the necessary knowledge for understanding the data types and structures, or statistical properties of the data. In many cases, users need to know how to get their data into a specific visualization tool as a first step. This relates to a more technical level of knowledge about a particular file format (e.g., CSV, JSON) or structure of the data – *data format* – (e.g., order and data types of individual variables).

**Visual encoding knowledge**: This type of knowledge is the most obvious one in the context of visualization, as it concerns the visual appearance of the data. Data elements are mapped to visual marks and channels to form visualizations. Understanding this mapping is the basis for being able to correctly interpret the visualization.

**Interaction knowledge**: Interactivity is crucial for visualization tools. An interactive visualization tool can support the investigation at multiple levels of detail, such as either a high-level overview or fully detailed views that show a small data subset only [44]. Understanding the interaction concepts used in a visualization tool is important for users for an active discourse with the data, i.e., to perform tasks and understand connections and relationships in the data.

**Analytical knowledge** is defined as the knowledge of different automated data analysis methods, for example clustering (e.g., k-means) or data aggregation (e.g., dimensionality reduction). In certain cases, users need to have at least a basic understanding of their characteristics in order to choose or parameterize them correctly.

### 3.2.2 HOW is visualization onboarding provided?

*Onboarding type, medium, context sensitivity, interaction, tool-specific, and educational theory* are relevant aspects of the question of how visualization onboarding is provided. The *onboarding type* captures the used *medium*. The form of *contextual aid* is extremely important for applications [22]. The help system should be designed to guide users by demonstration in the *context* of their own interface. Chilana et al. [11] developed an approach to provide a new framework for integrating crowd-sourced contextual help into web applications. In their work, they also discussed the importance of contextual help and *adaptive help* systems. Based on these results, we also integrate the aspect of *context sensitivity* into our framework for visualization onboarding. Fernquist et al. [17] introduced a set of the most relevant aspects for interactive tutorials for a sketching software. Based on their design space for sketching software, we adopted the aspect of *interactivity*. Additionally, we integrated the category *tool-specific* indicating if the onboarding concept is connected to a visualization tool or not. Visualization onboarding supports users in learning new concepts. Hence, we integrated the category of educational theory.

**Onboarding type & medium:** Onboarding can be provided in different types, such as guided tours, step-by-step wizards, video-based tutorials, and help centers. We derived this terminology from our literature review and Pronovic's blog article about context-sensitive and embedded help formats [12]. A particular type of onboarding consists of a *medium* which can be e.g., textual instructions, video, illustrations/figures, animations, etc.

**Context Sensitivity:** Context-sensitive help provides assistance at a specific point in the current state of the tool. It is the smallest possible chunk of information the user needs to understand at that point. Examples are in application help centers, guided tours, or mouseover popups including instructional material. A type of context-sensitive help is **embedded help** which goes beyond basic information and explanations by either detecting the user's need for help or offering a guided tour right on the interface. Examples are tooltips, instructions on the interface, or walkthroughs. **Context-free help** can be called at any state of usage and does not relate to the current state of help-seeking. Examples are online documentations and help videos.

**Interaction:** Interaction is applied within the onboarding process itself. We refer to Fernquist [17] for defining the degree of interactivity in onboarding concepts. Help systems can be **passive** if the user only consumes the learning material, such as reading an article or viewing a video. If users can try out the concepts, the onboarding concept is defined as **active**. Active tutorials that are aware of the users' interactions and can respond to these are referred to as **reactive**.

**Educational theory:** The aspect of learning and educational theories is crucial when it comes to onboarding approaches. A systematic categorization of the educational theories was not possible to conduct as there is no taxonomy available. Therefore, we collect educational theories, which authors described in their scientific publications (e.g., concreteness fading [2], learning by analog [52], etc)

**Tool-specific:** The category describes if the onboarding concept is designed for a specific visualization tool (tool-specific) or it is decoupled from it (non tool-specific).

### 3.3 WHERE is visualization onboarding provided?

Based on Fernquist et al. [17] who introduced a set of the most relevant aspects, we also adopted the aspect of the integration of onboarding concepts by asking *Where is visualization onboarding provided?* – externally, internally, or as a learning environment. An onboarding system that is integrated **internally** into the visualization can be more helpful for users because they do not have to jump back and forth between two different systems. **External** sources for onboarding concepts can be defined as sources which can be reached independently of the current state of the tool. At the tightest level of integration, help systems can be provided **internally**. It should be pointed out, however, that integrating onboarding systems into the visualization or visual analytics tools is challenging and requires a considerable effort.

### 3.4 WHEN is visualization onboarding used?

The aspect of *WHEN* describes the temporal aspect of intended onboarding use (see Figure 2 (6)). Onboarding concepts can be integrated **before** using the actual visualization tool (one time or repeated) or called up **while** the use of a certain tool, e.g., when support regarding a particular feature is needed.

# 4 Survey on Visualization Onboarding

In this section, we describe the method used for our systematic literature review in detail. Furthermore, we present the results of the survey based on our descriptive design space.

## 4.1 Method

To get a comprehensive overview of existing onboarding concepts, we systematically surveyed the literature published in the main venues in the fields of information visualization, visual analytics, and HCI. In addition to scientific publications, we reviewed commercial visual analytic tools based on a recent study about commercial systems by Behrisch et al. [6] (see Table 1). We focused on the following major conferences and journals: *IEEE InfoVis*, *IEEE VAST*, *EuroVis*, *Eurographics*, *EuroVA*, *IEEE TVCG*, *Information Visualisation* (IV), *ACM CHI*, and *ACM UIST*. Due to the fact that the term *onboarding* is rarely used in the visualization community, we used the following keywords: *data visualization literacy*, *visualization literacy*, *instructional material*, *learning*. We scanned the title and abstract for the specific keywords.

We additionally examined papers published as part of various relevant workshops on the topic of *visualization literacy*, especially the *IEEE VIS DECISIVE Workshop*. We took into account both full and short papers. Moreover, we identified the authors of the most relevant papers and included further publications by these researchers. We scanned through the related work sections of the relevant papers to find more literature related to our topic. We were able to identify a total of nine papers that focus on onboarding concepts and learning environments for visualization or visualization tools [2, 45, 40, 69, 62, 52, 41, 33, 8, 19] as well as ten commercial tools that use a variety of onboarding methods and concepts [54, 1, 53, 31, 63, 43, 55, 61, 64, 49].

Every selected publication was categorized by two coders who are co-authors of this paper. After the coding of the nine papers, we discussed the coding criteria and matched our coding strategy. In case of conflicting codes, coders discussed the reasons for decisions in order to resolve inconsistencies.

## 4.2 Results

We reviewed nine scientific publications and ten commercial tools with a special focus on onboarding concepts summarized in Table 1. In the following sections, we discuss and highlight the most relevant factors of onboarding methods we discovered.

| Name | Who? | | | | How? | | | | | | Where? | When? |
|---|---|---|---|---|---|---|---|---|---|---|---|---|
| | Domain Knowledge | Data Knowledge | Visual Encoding & Interaction Knowledge | Analytical Knowledge | Type | Medium | Tool-specific | Context-sensitivity | Interactivity | Educational Theory | external / internal | before / while |
| Alper et al., 2017 [3] | ● | ● | ● | ● | teaching tool | text, pictograms | — | context-free | active | concreteness fading | external | before, while |
| Ola & Sedig, 2017 [45] | n.a. | n.a. | n.a. | n.a. | video-tutorial | video | ✓ | context-free | passive | n.a. | external | before |
| Kwon & Lee, 2016 [40] | ○ | ○ | ● | ● | interactive tutorial walkthrough, video, static | screenshots, video, text, visual elements | — | context-free | active, passive, reactive | experimental learning model | external | before, while |
| Yalçın, 2016 [69] | ● | ● | ● | ○ | topic listing, point & learn, guided tour, notifications, topic answers | text | ✓ | context-sensitive, embedded | reactive | n.a. | internal | while |
| Tanahashi et al., 2016 [62] | ○ | ● | ● | ○ | InfoVis guide | text-plus-questions | — | context-free | active | top-down & bottom-up | external | before, while |
| Ruchikachorn & Mueller, 2015 [52] | ○ | ○ | ● | ○ | video tutorial | animated visualization sequences | — | context-free | active, passive | learning-by-analogy | external | before, while |
| Kang et al., 2003 [33] | ○ | ● | ● | ○ | step-by-step overlays | text | ✓ | context-sensitive, embedded | reactive | n.a. | internal | while |
| Bishop et al., 2020 [8] | ○ | ○ | ● | ○ | free-form constructive visualization tool | visual elements | — | context-free | reactive | scaffolding via visual feedback, learning from shared experiences | external | before, while |
| Firat et al., 2020 [19] | ○ | ● | ● | ○ | instructional software tool | visual elements, text | — | context-free | active | active learning | external | before, while |
| SAS *JMP* [54] | ○ | ● | ● | ● | website, video, step-by-step overlays | text, videos, images, visual elements | ✓ | context-sensitive, -free | active, passive, reactive | n.a. | internal, external | before, while |
| Advizor Solutions *Advizor* [1] | ○ | ● | ● | ● | website, overlay | text, videos, images | ✓ | context-sensitive, -free | passive | n.a. | external | before, while |
| SAP *Lumira* [53] | ○ | ● | ● | ● | website, video | text, videos, images | ✓ | context-free | passive | n.a. | external | before, while |
| IBM *Cognos Analytics* [31] | ○ | ● | ● | ○ | website, video, interactive guided tour | text, videos, images, visual elements | ✓ | context-sensitive, -free | active, passive, reactive | n.a. | internal, external | before, while |
| TIBCO *Jaspersoft* [63] | ○ | ● | ● | ○ | website, overlay | text, videos, images | ✓ | context-sensitive, -free | passive | n.a. | internal, external | before, while |
| Microsoft *Power BI* [43] | ○ | ● | ● | ○ | website, video, examples, ask questions in app | text, videos, images | ✓ | context-sensitive, -free | passive | n.a. | external | before, while |
| SAS *Visual Analytics* [55] | ○ | ● | ● | ● | website, video, images, interactive guided tour, courses, books | text, videos, images, visual elements | ✓ | context-sensitive, -free | active, passive | n.a. | internal, external | before, while |
| Tableau Software *Tableau* [61] | ○ | ● | ● | ● | website, video, courses, books | text, videos, images | ✓ | context-free | passive | n.a. | external | before, while |
| TIBCO *Spotfire* [64] | ○ | ○ | ● | ● | website, video, courses, books | text, videos, images | ✓ | context-free | passive | n.a. | external | before, while |
| QlikTech *QlikView* [49] | ○ | ● | ● | ○ | website, videos, courses, overlays | text, videos, images | ✓ | context-sensitive, -free | passive | n.a. | external | before, while |

**Table 1** Overview of available visualization onboarding approaches (rows), systematically characterized along the aspects of our conceptual framework (columns). The table is divided into academic research and concepts (upper half) and commercial tools (lower half) which make use of various onboarding concepts. With a main focus on the questions, we took four of them and mapped them to the categorization of available approaches. The colors refer to the equivalent questions explained in the subsections of 3. ● – applicable ○ – not applicable and *n.a.*: not available/unknown

#### 4.2.1 WHO – Who is the user? Which knowledge gap does the user have?

For both the scientific publications and the commercial tools, we recognized strong emphasis on *visual encoding and interaction knowledge* as well as *data knowledge* [2, 45, 40, 69, 62, 52, 41, 33, 54, 1, 53, 63, 43, 55, 49, 8]. Interestingly, Kwon and Lee [40], Ruchikachorn and Mueller [52], Bishop et al. [8], and the two visualization tools *IBM Cognos Analytics* [31] and *TIBCO Spotfire* [64] do not target *data knowledge explicitly*, which appears to be surprising as basic data knowledge is crucial in order to understand the visual encoding of a visualization. Only two publications [2, 40] cover *analytical knowledge*, while six of ten commercial tools provide support in this respect, e.g., classification and regression models [1]. We were able to identify a lack of *domain knowledge* in all tools and the majority of scientific publications. Only two publications focus on *domain knowledge* in their onboarding concepts [2, 69]. The publication of Ola and Sedig [45] was an exception insofar as we could not identify any of the knowledge gaps.

#### 4.2.2 HOW – How is visualization onboarding provided?

In this dimension, we distinguish between five different aspects: *onboarding type and medium, context-sensitivity, interactivity, tool-specific, educational theory* (see Section 3.2.2). In terms of the onboarding type and medium we found some similarities within the collection of publications. However, these have been the most difficult to gather, as the publications vary the most in their onboarding approaches. In the educational setting [2, 19, 8], the teaching tools use *text*, *visual elements*, as well as pictograms as *medium* to educate students. In terms of documented *onboarding type*, Alper et al. [2] introduced a "tool for teaching bar charts". More recently, Firat et al. [19] developed an instructional software tool for treemap visualizations, and Bishop et al. [8] introduced a "free-form constructive visualization tool". Besides, Kang et al. [34] as well as Yalçın [69] only integrated *text* in their onboarding approaches on *overlays*. Kang et al. [34] focused their concept on *step-by-step overlays*, in contrast, Yalcin [69] used for his approach overlays including a combination of *topic listing, point & learn, guided tour, notification, and topic answers*. A further similarity is the usage of video and/or animation to onboard users. For example, Ola and Sedig [45] as well as Ruchikachorn and Mueller [52] developed *video-tutorials* using *animated visualization sequences* [52] (see Figure 3 (3)) and a *video* [45] to support users in learning. In addition, we identified other types such as interactive walkthrough tutorials [40], and InfoVis Guides using text-plus-questions [62]. In general, most of the collected onboarding approaches use a combination of different medium and onboarding approaches.

All commercial tool could be systematically categorized only in terms of type and media using documentation/explanation websites with *screenshots* and *textual descriptions* (medium). The majority of tools also use *videos* as a medium to onboard users. *SAS JMP* [54], *IBM Cognos Analytics* [31] and *SAS Visual Analytics* [55] integrate step-by-step tutorials or interactive guided tours and therefore also rely

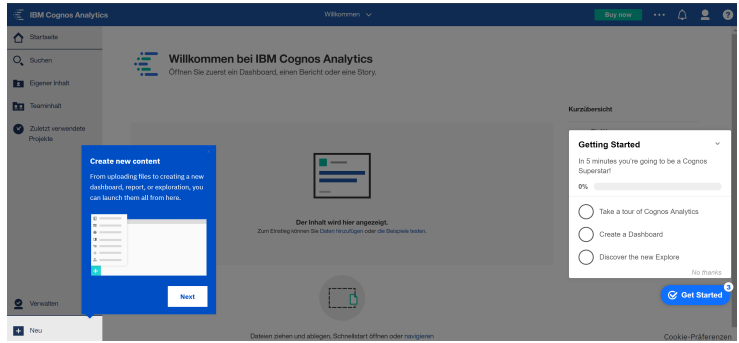
*(1) Commercial Visualization Tool IBM Cognos showing a guided tour*

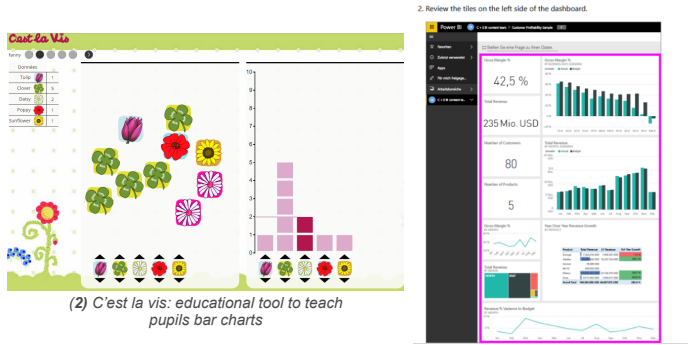
*(2) C'est la vis: educational tool to teach pupils bar charts*

*(3) Power BI Onboarding Page: Screenshots and text*

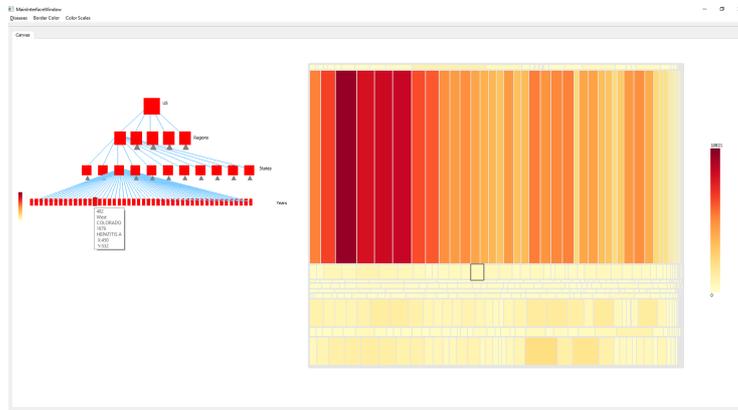
*(4) Instructional treemap tool interface with traditional tree structure (left) and linked treemap visualization (right).*

**Fig. 3** Onboarding approaches (1) IBM Cognos [31], (2) Alper et al. [2] onboarding method based on the concretness fading educational theory, (3) PowerBI external webpage with instructional material (screenshots and text) [43], and, (4) Educational instructional material for treemap visualization [19]

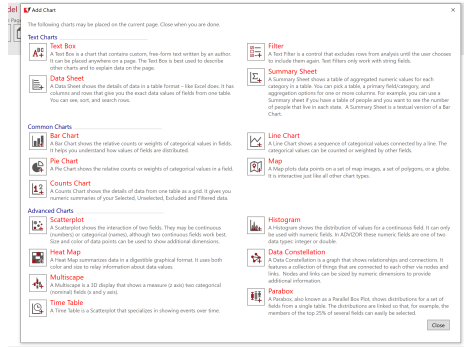

*(1) Commercial Visualization Tool Advizor showing textual instruction to the use of various visualization techniques*

*(2) Learning-by-Analogy: In-betweens of linear chart and spiral chart*

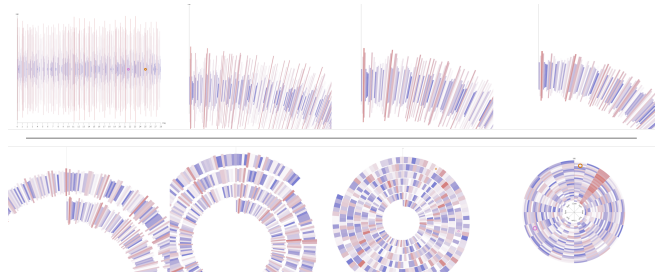

*(3) Onboarding for parallel coordinates*

**Fig. 4** Onboarding approaches (1) Advizor [1], (2) Learning by Analogy developed by [52], (3) Interactive tutorial based on Experiential Learning Model [40]

on visual elements (chart parts to interact with, applicable filters, etc.). *TIBCO Jaspersoft* [63] and *Advizor* [1] make use of an in-application help overlay using text and videos. Additionally *Microsoft Power BI* [43], *SAS Visual Analytics* [55], *Tableau* [61], *TIBCO Spotfire* [64], and *QlikTech QlikView* [49] provide a combination of books and courses. One special method to highlight is the in-application *ask questions* of *Microsoft Power BI* [43] which allows the users to ask a question related to the data set they are currently working on.

*Tool-specific:* For the scientific publications we identified three onboarding approaches which can be categorized as tool-specific [45, 69, 32]. The remaining six are non-tool-specific [2, 40, 62, 52, 8, 19]. We call these onboarding concepts *learning environments*, which are independent of a specific visualization tool and can be used in general.

*Context sensitivity* refers to the three categories context-free, context-sensitive, and embedded concepts. Seven out of nine papers designed context-free onbording concepts, while only Yalcin [69] and Kang et al. [33] use context-sensitive and embedded onboarding methods. On the other hand, three out of ten commercial tools integrate context-free onboarding concepts. The other commercial tools integrate context-free and context-sensitive methods as they are using documentation websites and also in-application overlays or guided tours. One example is *Advizor* [1] which makes use of context-free and context-sensitive onboarding methods (see Figure 3 (2) for the design of the context-sensitive approach).

A more detailed investigation of the *interactivity* of the onboarding concepts described in publications we found, revealed a good balance between the three types of interaction. The category *interactivity* is also connected with the used educational theory. Four of the nine onboarding concepts provide *reactive* onboarding [40, 69, 34, 8]. For the commercial tools, we observed a strong trend towards passive interactivity. Only two tools—*SAS JMP and IBM Cognos Analytics* [54, 31]— cover all three interactivity types. *IBM Cognos Analytics*, for example, provides videos and a website (passive) as well as an interactive guided tour (reactive) to onboard users.

In terms of the integrated *educational theories* we could not find any unique use of educational theories among the onboarding approaches presented in publications. Thus, we identified the following aspects: (1) *onboarding approach designed without the integration of educational theories* [45, 69, 34]; and (2) *onboarding approaches grounded in educational theories*: (2a) concreteness fading [2], (2b) Experiential learning model [40], (2c) top-down & bottom-up [62], (2d) learning-by-analogy [52], (2e) scaffolding via visual feedback, learning from shared experience [8], and (2f) active learning [19]. In the following, we describe two examples in detail showing how visualization onboarding has been applied.

*Example on Experiential Learning Model (2b):* One example for a reactive onboarding is from Kwon and Lee [40], who developed an online learning approach for parallel coordinates following the experiential learning model (see Figure 3 (6)). The model defines learning as the process in which knowledge is constructed via concrete experience and reflection on the experience [38]. Therefore, the interac-tive tutorial page integrates the experiential learning model's four-stages (Concrete Experience, Reflective Observation, Abstract Conceptualization, and Active Experimentation)). The authors implemented the model as follows. For the first stage, the *Concrete Experience*, the people are asked to complete a mission. For the *Reflection Observation* stage the onboarding approach provides hints to the user interactions. Additionally, "the system shows the conceptual goal of the the activity at a successful completion" [40] (*Abstract Conceptualization*). For the fourth level – *Active Experimentation* – the learning approach suggests to repeat the activity to strengthen the learning. They conducted a comparative evaluation with three tutorial types (static, video-based, and interactive tutorial walktrough). They observed that participants using the interactive and video tutorials outperformed participants with static or no tutorials.

*Example on Learning-by-Analogy (2d):* In addition to onboarding using the experiential learning model [40], Ruchikachorn and Mueller [52] proposed a concept for the teaching of unfamiliar visualizations by using the educational theory of *learning-by-analogy*. This is a example of a combination of *passive* and *active* onboarding system. Based on animated visualization sequences (passive) the users where taught a more advanced visualization technique based on an easier one with transitions as presented in Figure 3 (3). The user was able to watch the sequences which can be categorized as an *passive* interaction. Additionally, the user was able to start and stop animating the morphing (*active*).

### 4.2.3 WHERE – Where is visualization onboarding provided?

Our survey of existing work and commerical tools showed that the majority of onboarding solutions can either be classified as external or internal, or a combination of both sources. Yalçın [69] and Kang et al. [33] designed an internal onboarding concept. All other solutions can be categorized as external onboarding approaches. For commercial tools there is a fairly equal distribution between only *external* ones and those who are *external and internal*. The majority of commercial tools provide external material such as documentation sites with text, images, and videos.

### 4.2.4 WHEN – When is visualization onboarding used?

Onboarding concepts can be integrated at different states of use – before or during. Ola & Sedig [45] relied on a *before* approach, in contrast, Yalçın [69] and Kang et al. [33] provide their onboarding *while* the usage. Other onboarding approaches [2, 40, 62, 52, 8, 19] can be either used *before* or *while*. We detected a clear tendency for commercial tools as all of the onboarding concepts can be used before and during usage of the particular visualization tool.

### 4.3 Summary

Considering the **WHO** question we observed a strong tendency towards *visual encoding & interaction knowledge* [2, 40, 69, 62, 52, 33, 8, 19]. *Data knowledge* is also prominent in the literature [2, 69, 62, 33, 19]. However, *domain knowledge* [2, 69] and *analytical knowledge* [2, 40] are covered only by two out of nine investigated papers. Only Alper et al. [2] are targeting all knowledge gaps. Regarding the question of **HOW** is onboarding provided? — we found a variety of different onboarding types. This ranges from simple texts instructions [69] or videos [52, 45] to interactive visual elements [40, 2, 19, 8] or step-by-step guides [33]. Regarding context sensitivity, most of them are using a context-free approach [2, 45, 40, 62, 52, 8, 19], with two exceptions that are context-sensitive and embedded in the visualization tool [69, 33]. Those two exceptions are also *internal* looking at the **WHERE** aspect. All others are designed as non tool-specific onboarding approaches, i.e., not directly integrated into a visualization tool which are then *external*.

In the case of *educational theory*, however, no general statement can be made based on the categorization of the papers, since each paper follows a different educational theory. However, we observed similarities regarding the educational theories, which are presented in Section 4.4. In general, most of the collected onboarding approaches of the commercial tools are design to be used *before* and *while* interacting with a particular visualization tool (**WHEN**).

None of the commercial tools address or attempt to explain the *domain knowledge* of the users. The tools mainly cover only the *data knowledge* [54, 1, 53, 63, 43, 55, 61, 49] as well as the *visual encoding & interaction knowledge* [54, 1, 53, 31, 63, 43, 55, 61, 64, 49]. In general, the tendency to convey analytical knowledge is much higher with commercial tools [54, 1, 53, 55, 61, 64] than with the scientific papers. In relation to tools, the type of **onboarding** mainly relies on help websites, video tutorials or courses. There are a few exceptions [54, 31, 55] that also use visual elements offering more interaction. For *context-sensitivity*, it is about evenly distributed among the tools, but there is no single embedded one. Also the *interactivity* in the tools is mostly *passive* since the help often is only provided on demand. Exceptions to this are the three approaches [54, 31, 55] that offer guides or tutorials directly or react to user interaction. Unfortunately, it was not possible to identify an *educational theory* for any of the commercial tools, but this was to be expected, since they are established visualization software. The commercial tools have a balanced ratio in the question of **WHERE**. In terms of the **WHEN**-question, all the onboarding approaches can be used while or before using the actual visualization tool.

### 4.4 Existing Design Considerations for Visualization Onboarding

In this section, we present existing guidelines derived from the collected papers. We focused on the given medium, type of onboarding, as well as the education theory used to onboard users.

- Kwon and Lee [40] developed an interactive guide for parallel coordinates plots based on a **learning-by-doing** approach. They followed the "Experiential learning model", which can be defined as the process in which knowledge is constructed via concrete experience and reflection on the experience [38]. The presented interactive tutorial walkthrough integrates textual descriptions as well as interactive visual elements (see Figure 4–(3)), where, for example, the user can click on points in integrated parallel coordinates, whereupon lines are drawn that then connect them.
- In their paper, Ruchikachorn and Mueller [52] developed a teaching concept to learn and teach unfamiliar visualizations by linking it to a more familiar one. They followed the **learning-by-analogy** approach. The authors commented that their system can be useful when the visualization method to be learned is inherently more powerful than its counterpart. Their approach overcomes languages barriers as it uses visuals.
- The results of the conducted study by Tanahashi et al. [62] showed that tutorials where users can directly interact with the visualization will influence the comprehension positively. They suggest to use **active learning** type (participating actively in a corresponding dialog) with **top-down** exercises. In detail, this means to ask participants to draw more advances, less direct inferences from the data. Their study revealed that their approach of text-plus-question introductory tutorials is a useful and practical way to onboarding users to information visualizations.
- A recent study shows that there is a successful knowledge transfer to another concrete domain when concrete examples were given as opposed to abstract ones [13]. Based on this results Alper et al. [2] developed a tablet app teaching elementary school pupils bar charts using the pedagogical method of **concreteness fading**. The tool provides a space with a reference line (x- and y-axis) as well as free-form pictograph that represents data in the form of illustrative icons which are scattered around. Children can stack the icons on top of each other and then watch an animated transitions morphing the icons into a more abstract representations of a bar chart (see Figure 3 (2)).
- Bishop et al. [8] developed a free-form construction tool for tablets to engage pupils with the creation of visualization, as well as to make the visual mapping of data more explicit. **Scaffolding** was integrated as educational theory. The results of their study highlight the advantage of scaffolding within the creation process of visualizations through visual feedback, configurability, and shared interaction.

When we sum up and generalize the results of the empirical studies of the papers, as well as the results of the analysis of the design space, we propose the following guidelines when it comes to design onboarding methods: (1) explain the visual encoding & interaction concepts [2, 8, 40]; (2) use interactive onboarding approaches, where users can interact with the visualization as well as with the instructional material [62, 40]; (3) concrete experience and reflection can lead to higher understanding [40, 62]; (4) use animations or videos [2, 52, 40] to show the data-to-visual mapping.

## 5 Discussion & Conclusion

We presented a descriptive design space for visualization onboarding and presented design considerations based on the existing empirical studies. The design space contains the six aspects: **WHY** is visualization onboarding needed? **WHAT** is visualization onboarding? **WHO** is the user? Which knowledge gap does the user have? **HOW** is visualization onboarding provided? **WHERE** is visualization onboarding provided? **WHEN** is visualization onboarding used? We conducted a systematic literature review to develop the presented design space. Additionally, we also reviewed commercial visualization tools listed in Table 1. We especially focused on educational theories as the aspect of learning is important when it comes to the design of visualization onboarding (see Table 1 and Section 4.4). Ways to effectively support the learning process of users with different knowledge gaps can be considered by using educational theories. However, the literature lacks educational theories with a special focus on onboarding concepts. We tried to identify guidelines based on the existing literature, which we presented in Section 4.4. Nevertheless, existing theories and results of educational research can be used to inform the design of onboarding systems.

Onboarding systems can either be designed like help systems, which implies a cognitivist approach, or they might use a scaffolding approach [8], applying features such as prompts, tools to structure information or higher-order questions. Constructivist theory supports the assumption that especially higher-order reasoning processes and the ability to make inferences and draw conclusions from the data are supported by cognitive apprenticeship or scaffolding in particular. Higher-order reasoning is the last stage in the model suggested by graph comprehension but also the ultimate goal of most visualization systems. Based on the papers, educational theories which support active learning and concrete experience are appropriate for onboarding. Further research is needed to empirically test these observations.

## Acknowledgments

This work was funded by the Austrian Science Fund as part of the VisOnFire project (FWF P27975-NBL), the Austrian Ministry for Transport, Innovation and Technology (BMVIT) under the ICT of the Future program via the SEVA project (no. 874018), as well as the FFG, Contract No. 854184: "Pro2Future", which is funded within the Austrian COMET Program Competence Centers for Excellent Technologies under the auspices of the Austrian Federal Ministry for Transport, Innovation and Technology, the Austrian Federal Ministry for Digital and Economic Affairs and of the Provinces of Upper Austria and Styria. COMET is managed by the Austrian Research Promotion Agency FFG.